\documentstyle[twoside,fleqn,epsf,espcrc2]{article}

\title{The domain wall fermion chiral condensate in quenched QCD}

\author{P.~Chen\address{Department of Physics, Columbia University, 
New York, NY 10027}\thanks{Research supported in part by the U.S. Dept.
of Energy.},
N.~Christ$^{\rm a\;*}$,
G.~Fleming\thanks{Talk presented by G.~Fleming.}$^{\rm a\;*}$,
A.~Kaehler$^{\rm a\;*}$,
C.~Malureanu$^{\rm a\;*}$,
R.~Mawhinney$^{\rm a\;*}$,
G.~Siegert\thanks{Supported by the Max Kade Foundation},
C.~Sui$^{\rm a\;*}$,
P.~Vranas
\thanks{Currently at Physics Dept., 
University of Illinois, Urbana, IL 61801.}$^{\rm a\;*}$, and
Y.~Zhestkov$^{\rm a\;*}$
}

\begin{document}

\def\thepage{CU--TP--916}

\begin{abstract}
We examine the chiral limit of domain wall fermions in quenched QCD.
One expects that in a quenched simulation, exact fermion zero modes
will give a divergent, $1/m$ behavior in the chiral condensate for
sufficiently small valence quark masses. Unlike other fermion
formulations, domain wall fermions clearly demonstrate this behavior.

\end{abstract} 

\maketitle

\section{INTRODUCTION}
\label{sec:intro}

It is a common assertion that quenched lattice QCD qualitatively
produces many of the features found in continuum full QCD.  However, a
quenched simulation is expected to include topological gauge
configurations which produce exact fermion zero modes.  Not suppressed
by the missing fermion determinant, such modes will cause divergent
behavior in the small mass limit of the chiral condensate.

Surprisingly, the standard lattice fermion formulations do not
demonstrate divergent behavior for the quenched chiral condensate at
current lattice spacings and quark masses.  One may imagine that the
small eigenvalue spectrum of those lattice Dirac operators bears
little resemblance to the continuum Dirac spectrum until the lattice
spacing is substantially smaller. See the introduction of
\cite{zero_modes} for further discussion and references.

Starting from Kaplan's initial ideas for domain wall fermions (DWF)
\cite{kaplan}, several fermion formulations have been developed which
use an infinite number of massive fermion flavors to produce an
effective chiral-invariant action for a single massless fermion
flavor.  Using the overlap formalism \cite{nn95}, which describes the
infinite flavor case, Narayanan and Neuberger demonstrated that these
theories have a unique, integer valued index for the associated Dirac
operator.  Studies have shown that nearly exact zero modes persist in
truncated theories, provided the number of heavy flavors is large
enough. One should then expect to see these zero mode effects in
quenched simulations.

\section{THE CONTINUUM PICTURE}
\label{sec:continuum}

Using a spectral decomposition of the continuum
Dirac operator, we can express the quark chiral condensate as a function
of the quark mass for any ensemble of gauge fields
\begin{eqnarray}
\left\langle\bar\psi\psi\right\rangle&=&\frac{1}{V}\int d^4x 
\left\langle\bar\psi(x)\psi(x)\right\rangle\nonumber\\*
\label{eq:pbp}
&=& \frac{N_{\rm zm}}{m} + 2 m \int_0^\infty
d\lambda \frac{\rho(\lambda)}{\lambda^2+m^2},
\end{eqnarray}
where we have used the $\gamma_5$ symmetry of the Dirac spectrum and 
$N_{\rm zm}$ are the number of zero eigenvalues in the spectrum.
We can similarly express the integrated pion correlator
\begin{equation}
\label{eq:pi_corr}
{1 \over V}\int d^4x \left\langle\pi(x)\pi(0)\right\rangle = \frac{N_{\rm zm}}{m^2}
+ 2 \int_0^\infty d\lambda \frac{\rho(\lambda)}{\lambda^2+m^2}.
\end{equation}
Spontaneous chiral symmetry breaking is related to a non-zero
eigenvalue density at $\lambda=0$ by the Banks--Casher relation which,
for the case of non-zero $N_{\rm zm}$, becomes

\begin{equation}
\label{eq:banks_casher}
\left\langle\bar\psi\psi\right\rangle
\sim \frac{N_{\rm zm}}{m} + \pi\rho(0) + {\cal O}(m), \quad m\to 0.
\end{equation}
Finally, we note that the identity 
\begin{equation}
\label{eq:m_pi_corr_o_pbp}
m \int d^4x \left\langle\pi(x)\pi(0)\right\rangle
= \left\langle\bar\psi\psi\right\rangle
\end{equation}
follows from the continuum chiral symmetry as represented by
equations (\ref{eq:pbp}) and (\ref{eq:pi_corr}).

\def\thepage{\arabic{page}}

\section{DOMAIN WALL FERMIONS}
\label{sec:dwf}

Our numerical simulations were performed using the boundary fermion
variant of domain wall fermions \cite{Furman_Shamir}.  Here one adds a
fifth dimension to the normal 4-D space time of extent $L_s$.  The
gauge fields are not changed by this extension and do not depend on
$s$.  There are also two mass parameters for the fermions:  $m_f$, the
explicit four-dimensional bare quark mass and $m_0$, the
five-dimensional bare quark mass.  Except for different symbols for the
three parameters listed above, our conventions follow
\cite{Furman_Shamir}.

In the free field case, the effective mass of the low energy
state is \cite{PMV}:
\begin{equation}
\label{eq:free_m_eff}
m_{\rm eff}^{\rm (free)} = m_0(2-m_0) \left[m_f+(1-m_0)^{L_s}\right] .
\end{equation}
where, for one flavor physics,  $m_0$ should be in the range $(0,2)$.
This simple equation reveals two important features of domain wall
fermions.
First, as $L_s\to\infty$, $m_{\rm eff} \propto m_f$ and the
proportionality constant depends on the five--dimensional mass $m_0$.
Second, for finite $L_s$, the Wilson-like additive correction to
$m_{\rm eff}$ is exponentially suppressed in $L_s$ and the rate of
decay is determined again by $m_0$.  In the
interacting case, we expect that the effective mass will behave 
in a similar fashion 
where the residual mass term $m_{\rm res}$ should vanish exponentially in
$L_s$ at a rate governed by $m_0$ and the coupling constant. Indeed, 
numerical studies \cite{PMV} indicate that the
decay remains exponential with a decay rate that becomes faster as the
continuum limit is approached.  
\begin{figure}[htb]
\epsfysize=2.4in
\vspace{-0.5cm}
\epsfbox[43 43 522 452]{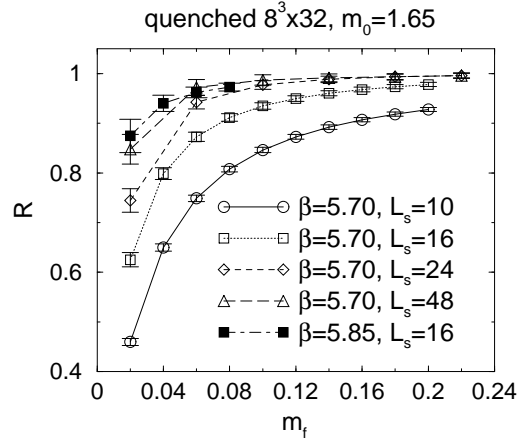}
\vspace{-1.3cm}
\caption{Ratio $R$ of $m \int d^4x \langle\pi(x)\pi(0)\rangle$ to $\langle\bar\psi\psi\rangle$}
\vspace{-1.0cm}\end{figure}

\section{QUENCHED DWF AT ZERO TEMPERATURE}
\label{sec:zero_temp}

We begin by examining the extent to which domain wall fermions
reproduce the effects of chiral symmetry for finite $L_s$ in a quenched
simulation by computing the ratio of the two sides of
(\ref{eq:m_pi_corr_o_pbp}) on $8^3 \times 32$ lattices at $\beta = 5.7$
and 5.85.  We choose to work at a domain wall height, $m_0=1.65$.  This
choice is supported by numerical studies of the chiral condensate for
several values of $m_0, L_s$ and $m_f$ at these values of $\beta$.

In Figure 1 we show the ratio of quantities proportional to the left and
right hand sides of (\ref{eq:m_pi_corr_o_pbp}).  The constant,
$m_f$-independent, behavior of this ratio, required by chiral symmetry,
is increasingly visible in Figure 1, for the curves with larger $L_s$.
Particularly striking is the comparison of this ratio between
$\beta=5.7$ and 5.85.  The weaker coupling with $L_s=16$ shows a degree
of chiral symmetry present at the stronger coupling only for a three
times larger $L_s$.  This is consistent with the picture that
stronger coupling requires larger $L_s$ \cite{PMV}.

\begin{figure}[htb]
\epsfysize=2.4in
\vspace{-0.5cm}
\epsfbox[16 19 546 476]{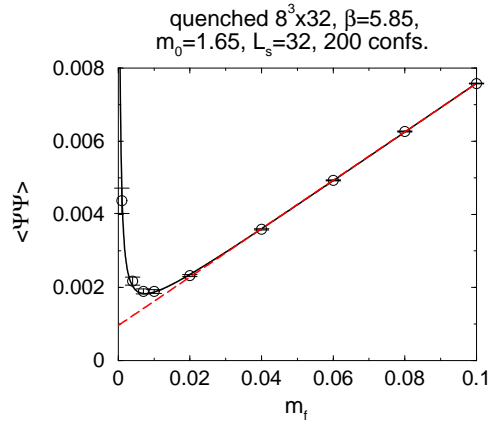}
\vspace{-1.5cm}
\caption{$1/m_f$ divergence in $\langle\bar\psi\psi\rangle$}
\vspace{-1.0cm}\end{figure}

Next we study the small mass limit of $\langle\bar\psi\psi\rangle$
computed on 200 quenched configurations at $\beta=5.85$ and $L_s=32$.
The results, shown in Figure 2, show a clear signal for a $1/m_f$
divergence for $m_f < 0.01$.  The solid line is a fit for $0.001 \le
m_f \le 0.1$ which includes a $1/m_f$ term and the dashed line is a
linear fit for $0.01 \le m_f \le 0.1$.  The constant and linear terms
are consistent between these two fits within errors and the
$\chi^2$/dof is one for both fits.  Table 1 summarizes the rest of our
$T=0$ results.  We have used $m_0=1.65$ for all simulations.

We believe that this is the first time that the $1/m_f$ behavior in
$\langle\bar\psi\psi\rangle$ expected from fermion zero modes present
in quenched calculations has been seen.  Notice the decrease in the
strength of this term, shown in Table 1, when the volume is increased
from $8^3\times 32$ to $16^3\times 32$.

\begin{table}[htb]
\caption{Fit to $\langle\bar\psi\psi\rangle=c_{-1}/m_f+c_0+c_1m_f$}
\begin{tabular}[ht]{llll}
\hline
                          & $8^3\times 32$ & $8^3\times 32$ & $16^3\times 32$ \\
                          & $\beta=5.70$   & $\beta=5.85$   & $\beta=5.85$    \\
                          & $L_s=48$       & $L_s=32$       & $L_s=32$        \\
                          & 40 configs     & 200 configs    & 90 configs      \\
\hline
$c_{-1} (\times 10^{-6})$ & 1.6(4)         & 3.8(3)         & 0.60(9)         \\
$c_0 (\times 10^{-3})$    & 2.15(3)        & 0.82(3)        & 0.89(1)         \\
$c_1 (\times 10^{-2})$    & 6.01(2)        & 6.73(3)        & 6.71(2)         \\
$\chi^2$/dof              & 8.99           & 1.24           & 2.96            \\
\hline
\end{tabular}
\end{table}

\normalsize

\section{QUENCHED DWF AT FINITE TEMPERATURE}
\label{sec:finite_temp}

\begin{figure}[htb]
\epsfysize=3.5in
\vspace{-0.5cm}
\epsfbox[48 33 583 650]{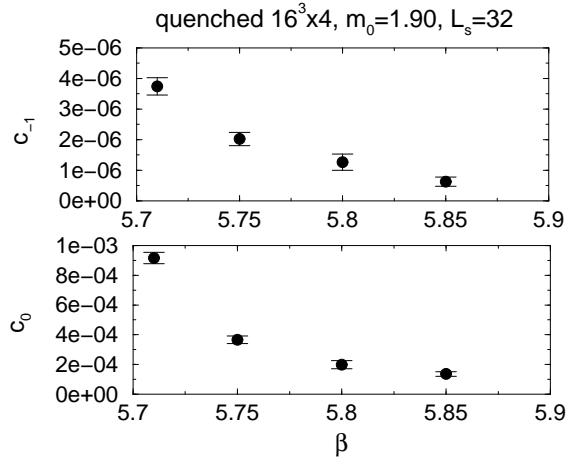}
\vspace{-1.0cm}
\caption{Fit coefficients for $T>T_c$}
\vspace{-0.5cm}
\end{figure}

In Table 2 we summarize a quenched study of
$\langle\bar\psi\psi\rangle$ on a $16^3\times 4$ lattice evaluated for
$\beta$ slightly above $\beta_c=5.6925$.  Note, we continue to see a
divergent behavior of $\langle\bar\psi\psi\rangle$.  To our surprize,
we also see a non-zero constant contribution to the chiral condensate
well into the symmetric phase.  Figure 3 shows the behavior of both
$c_{-1}$ and $c_0$ as we increase $\beta$ above $\beta_c$.  Each shows
the rapid decrease as the system heats up expected if their origin is
topology-induced fermionic, zero modes.  This never-before-seen,
quenched small-mass behavior of $\langle\bar\psi\psi\rangle$, suggests
that domain wall fermions offer new insights into the continuum
behavior of quenched QCD.

\begin{table}
\caption{$16^3 \times 4, m_0=1.90, L_s=32$}
\begin{tabular}[ht]{lllll}
\hline
             & $c_{-1}$           & $c_0$              & $c_1$              &              \\
$\beta$, conf & $(\times 10^{-6})$ & $(\times 10^{-3})$ & $(\times 10^{-2})$ & $\chi^2$/dof \\
\hline
5.71, 128     & 3.7(3)             & 0.92(4)            & 8.97(5)            & 5.34         \\
5.75, 116     & 2.0(2)             & 0.37(3)            & 9.14(3)            & 1.44         \\
5.80, 84      & 1.3(3)             & 0.20(3)            & 9.11(3)            & 0.30         \\
5.85, 136     & 0.6(1)             & 0.14(2)            & 9.02(2)            & 0.81 \\
\hline
\end{tabular}
\end{table}


\normalsize

\begin{thebibliography}{99}

\bibitem{zero_modes} P.~Chen, {\it et.~al.}, hep-lat 9807029.

\bibitem{kaplan} D.~B.~Kaplan, Phys.~Lett.~B 288 (1992) 342.

\bibitem{DWF_reviews} R. Narayanan,
Nucl. Phys. {\bf B34} (Proc. Suppl.)  (1994) 95;
M. Creutz, Nucl. Phys. {\bf B42} (Proc. Suppl.)  (1995) 56;
Y. Shamir, Nucl. Phys. {\bf B47} (Proc. Suppl.)  (1996) 212;
T. Blum, these proceedings.

\bibitem{nn95} R.~Narayanan and H.~Neuberger, Nucl.~Phys.~B 443 (1995) 305.

\bibitem{Furman_Shamir} V. Furman, Y. Shamir, Nucl. Phys. {\bf B439} (1995) 54.

\bibitem{PMV} P.M. Vranas, Nucl. Phys. {\bf B53} (Proc. Suppl.) (1997) 278;
Phys. Rev. {\bf D57} (1998) 1415.

\end{thebibliography}
\end{document}